\documentclass[floatfix,a4paper,tightenlines,amsmath,amssymb,nofootinbib,showpacs,notitlepage,showkeys,prd,twocolumn,10pt]{revtex4-1}

\usepackage[utf8]{inputenc}
\usepackage{graphicx}
\usepackage{hyperref}

\newcommand{\de}{\delta}
\newcommand{\eref}[1]{Eq.~(\ref{#1})}
\newcommand{\tref}[1]{Tab.~\ref{#1}}
\newcommand{\fref}[1]{Fig.~\ref{#1}}
\newcommand{\nnnl}{\nonumber\\}	

\newcommand{\MeV}{\text{MeV}}

\allowdisplaybreaks 

\begin{document}

\title{Phase structure and propagators at nonvanishing temperature for QCD and QCD-like theories}

\author{Romain Contant}
\email{romain.contant@uni-graz.at}
\affiliation{Institute of Physics, University of Graz, NAWI Graz, Universit\"atsplatz 5, 8010 Graz, Austria}

\author{Markus Q.~Huber}
\email{markus.huber@uni-graz.at}
\affiliation{Institute of Physics, University of Graz, NAWI Graz, Universit\"atsplatz 5, 8010 Graz, Austria}

\date{\today}

\begin{abstract}
We investigate the universality of truncation schemes for Dyson-Schwinger equations developed for quantum chromodynamics in theories which differ from quantum chromodynamics only in the gauge group. Our specific choices are the gauge groups $SU(2)$ and $G_2$, for which lattice calculations at nonvanishing chemical potential are possible. Thus, corresponding calculations can provide benchmarks for testing calculations with functional equations. We calculate the quark and gluon propagators and determine the chiral and dual chiral condensates at vanishing density to determine the confinement/deconfinement and chiral transitions, respectively.
We can reproduce the expected type of transitions in the quenched and unquenched cases. In general, all three theories react very similarly to modifications of the employed model for the quark-gluon vertex.
\end{abstract}



\maketitle

\section{Introduction}

The phase diagram of quantum chromodynamics (QCD) contains a plethora of interesting physics. Correspondingly much effort is devoted to its investigation both with experiments and from the theory side. However, the phase structure at nonvanishing chemical potential, where a critical point and additional phases are expected, is still elusive, since we are lacking a method that is based on first principles and works reliably in that regime. In particular, the otherwise very successful method of Monte Carlo lattice simulations is plagued by the complex action problem that makes standard simulations at nonzero chemical potential unfeasible \cite{deForcrand:2010ys}. Many different approaches are being pursued to overcome this problem, e.g., \cite{Gattringer:2014nxa,Aarts:2015kea,Scorzato:2015qts,Sakai:2016tzv,Langfeld:2016kty}.

A method complementary to lattice simulations is functional equations like equations of motion of correlation functions \cite{Cornwall:1974vz,Roberts:1994dr,Alkofer:2000wg,Berges:2004pu,Alkofer:2008nt} and the functional renormalization group \cite{Berges:2000ew,Pawlowski:2005xe,Gies:2006wv}. Since they form infinitely large systems of equations, only a subset of equations can be solved. The choice of such a subset requires to specify how to treat the correlation functions not included in the subset. In the following we refer to the specification of which subset of equations is calculated \textit{and} the definitions of the missing correlation functions as truncation. The corresponding models are constrained by various requirements like their known asymptotic behavior or by results from other methods. A particularly useful source of information is results of correlation functions from lattice simulations. For example, quantitative results for the gluon propagator can be provided by lattice simulations and provide benchmarks for functional calculations.

The interplay between functional and lattice methods has led to the development of some useful hybrid methods. They exploit the fact that some objects can be obtained more easily or that systematic errors are better under control in one method than in the other. For example, vertices are still a challenge for the lattice approach. The amount of statistics required is limiting to some extent and typically only restricted kinematic configurations are investigated. On the other hand, the quenched vacuum gluon propagator is by now well studied on the lattice \cite{Cucchieri:2007md,Cucchieri:2008fc,Sternbeck:2007ug,Bogolubsky:2009dc,Maas:2011se,Oliveira:2012eh}, while for functional equations a large effort is required to obtain a quantitative description due to the coupling to higher correlation functions, see, e.g., \cite{Huber:2012kd,Aguilar:2013vaa,Blum:2014gna,Cyrol:2016tym,Huber:2016tvc,Cyrol:2017ewj}. Additionally, respecting gauge covariance is challenging \cite{Huber:2014tva,Cyrol:2016tym,Aguilar:2016vin}.

At nonzero temperature also lattice results for the quenched \cite{Cucchieri:2007ta,Fischer:2010fx,Maas:2011ez,Aouane:2011fv,Cucchieri:2012nx,Silva:2013maa,Silva:2016onh} and unquenched \cite{Furui:2006py,Bornyakov:2011jm,Aouane:2012bk} gluon (and ghost) propagators are available. These results can be used as input for functional equations to avoid the quantitative uncertainties (and technical complexities) when solving for the gluon propagator directly. Using an approximated equation for the gluon propagator, also unquenching effects can be included \cite{Nickel:2006vf,Fischer:2011mz,Fischer:2012vc}. With this method, the transitions between the quark-gluon plasma and the hadronic phases have been investigated in a series of papers for two flavors \cite{Fischer:2011mz,Fischer:2012vc}, three flavors \cite{Fischer:2012vc,Fischer:2014ata} and four flavors \cite{Fischer:2014ata}. First results of the effects of baryons were also obtained \cite{Eichmann:2015kfa}. In addition, this approach was applied to low temperatures and high densities \cite{Muller:2016fdr}. In that parameter space also the existence of inhomogeneous phases was investigated \cite{Muller:2013tya}. Alternatively, phenomenologically motivated effective interactions combining the gluon propagator and the quark-gluon vertex into a single quantity are also used to study the phase diagram of QCD, e.g., \cite{Gutierrez:2013sta,Shi:2014zpa,Xin:2014ela,Gao:2016qkh}.

A result of all these investigations was that the critical point, where the crossover turns into a first order transition line, is at a quark chemical potential larger than the temperature. This is in agreement with corresponding extrapolations from lattice results. However, in order to obtain reliable quantitative values for its location, we would need to know if the applied approximations are still well justified in this region of the phase diagram. For example, how well is the quark-gluon interaction modeled and what influence do hadronic degrees of freedom have?
To investigate this, we make a detour to QCD-like theories that do not suffer from a complex action problem. Thus, lattice results at nonvanishing chemical potential can serve as benchmarks for functional results. Specifically, we will investigate QC$_2$D and $G_2$-QCD which correspond to QCD with the gauge group $SU(3)$ replaced by the gauge groups $SU(2)$ \cite{Kogut:2000ek} and $G_2$ \cite{Holland:2003jy,Pepe:2006er}. These theories have been and are still investigated on the lattice, e.g.,  \cite{Hands:2006ve,Cossu:2007dk,Danzer:2008bk,Hands:2010gd,Maas:2012wr,Ilgenfritz:2012wg,Hands:2012yy,Boz:2013rca,Wellegehausen:2013cya,Bruno:2014rxa,Braguta:2016cpw,Holicki:2017psk,Wellegehausen:2017gba} and with continuum methods, e.g., \cite{Brauner:2009gu,Andersen:2010vu,Strodthoff:2011tz,vonSmekal:2012vx,Strodthoff:2013cua,Khan:2015puu}. A direct application of corresponding lattice results at nonzero chemical potential is, for example, the study of the mass-radius relation of neutron stars in $G_2$-QCD \cite{Hajizadeh:2017jsw}.

QC$_2$D and $G_2$-QCD have many properties in common with QCD. In particular, they are all three asymptotically free and they feature dynamical chiral symmetry breaking and confinement. The transitions related to the last two phenomena coincide in the quenched case \cite{Karsch:2003jg,Danzer:2008bk,Lucini:2012gg} and are at least close when they become smooth transitions for the unquenched case \cite{Borsanyi:2010bp,Bazavov:2011nk,Maas:2012wr}. This last property is not automatic, as the example of QCD with adjoint fermions shows \cite{Karsch:1998qj,Engels:2005rr,Bilgici:2009jy}. The similarities go even further and extend to the level of the underlying correlation functions. This is shown by lattice results for the $SU(2)$ Yang-Mills propagators in the vacuum \cite{Sternbeck:2007ug,Cucchieri:2007zm}, but also at nonvanishing temperature \cite{Fischer:2010fx,Maas:2011ez}. In the latter case, differences occur below the phase transition which reflect the different orders of the transitions in $SU(2)$ and $SU(3)$. For $G_2$ we know at least in the vacuum in two and three dimensions \cite{Maas:2007af} that the Yang-Mills propagators are qualitatively similar. In functional equations, the different gauge groups are reflected by different Casimir operators of the groups. Differences to QCD are, for example, an extended flavor symmetry in QC$_2$D, the so-called Pauli-G\"ursey symmetry, and the existence of diquarks for both $G_2$-QCD and QC$_2$D. Also, at nonvanishing density QC$_2$D has a transition from a Bose-Einstein condensate (BEC) to a Bardeen-Cooper-Schrieffer (BCS) phase.

As a first step towards comparisons between QCD and QCD-like theories, we investigate here the case of vanishing chemical potential to study how universal the applicability of the one and same truncation scheme for all three theories is. Using the chiral and dual chiral condensates to distinguish the hadronic and quark-gluon plasma phases, we will consider the quenched and unquenched situations and test the sensitivity of the systems on the modeled part of the input.
First results for $G_2$ have been presented in \cite{Contant:2016ndj}.

Our setup is detailed in Sec.~\ref{sec:setup}. The results are presented in Sec.~\ref{sec:results} and we summarize in Sec.~\ref{sec:summary}. Two appendices contain details on the fits for the $SU(2)$ gluon propagator and a study of the importance of the dressing function $D(\vec{p},\omega_n)$ of the quark propagator.

\section{Setup}
\label{sec:setup}

The truncated system of equations considered here consists of the Dyson-Schwinger equations (DSEs) of the quark and gluon propagators. We will discuss these equations in turn and then the model employed for the quark-gluon vertex. Finally, the definitions of the observables employed for distinguishing the phases are given.

\subsection{Quark propagator}

\begin{figure}[tb]
 \begin{center}
 \includegraphics[width=0.45\textwidth]{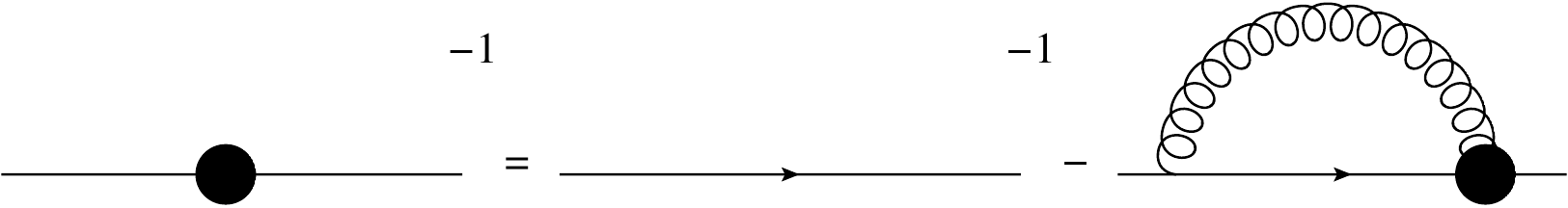}
 \caption{Quark propagator DSE. Quantities with a blob are fully dressed, as are internal propagators. Continuous/wiggly lines denote quarks/gluons.}
 \label{fig:DSEqqb}
 \end{center}
\end{figure}

At nonvanishing temperature the quark propagator has four components. Its inverse can be parametrized by
\begin{align}
S^{-1}(\vec{p}, \omega_{n}) &= i \vec{p} \vec{\gamma} A(\vec{p}, \omega_{n}) + i \omega_{n} \gamma_{4} C(\vec{p}, \omega_{n})\nnnl
&+ B(\vec{p}, \omega_{n}) +  i \omega_{n} \gamma_{4} \vec{p} \vec{\gamma} D(\vec{p}, \omega_{n}).
\end{align}
The dressing functions $A(\vec p, \omega_n)$, $B(\vec p, \omega_n)$, $C(\vec p, \omega_n)$, and $D(\vec p, \omega_n)$ contain the nonperturbative information. In the following we drop $D(\vec p, \omega_n)$. In various limits (vacuum, perturbation theory, chirally symmetric phase) it is zero. In Appendix \ref{sec:quarkD} we show explicit results for $D(\vec p, \omega_n)$ which confirm that for nonvanishing chemical potential $D(\vec p, \omega_n)$ is extremely small and thus irrelevant. The DSE of the quark propagator, diagrammatically shown in \fref{fig:DSEqqb}, reads
\begin{widetext}
\begin{align}\label{eq:quarkDSE}
S^{-1}(\vec{p}, \omega_{n})&=Z_2 S_0^{-1}(\vec{p}, \omega_{n}) -\Sigma(\vec p, \omega_n),\\
 \Sigma(\vec p, \omega_n)&= - Z_{1F} C_{F} g^{2}\sum_{q_4}{\int{\frac{d\vec{q}}{(2\pi)^3}\gamma_{\mu} S(q) \Gamma_{\nu}(p-q;-p,q) D_{\mu \nu}(p-q)}}.
\end{align}
\end{widetext}
$C_F$ is the Casimir of the gauge group from the color trace. $Z_2$ and $Z_{1F}$ are the quark wave function and the quark-gluon vertex renormalization constants, respectively. The bare quark propagator is given by
\begin{align}
S_0(\vec{p}, \omega_{n})=(i \vec{p} \vec{\gamma} + i \omega_{n} \gamma_{4} + m_0)^{-1},
\end{align}
where $m_0=Z_m m_R$ is the bare quark mass, $m_R$ the renormalized quark mass and $Z_m$ the quark mass renormalization constant. For brevity, we sometimes use the four-momentum, although the frequency has to be treated separately at nonvanishing temperature: Gluons have discrete Matsubara frequencies $p_4=\omega_n=2\pi\,n\,T$ and quarks $p_4=\omega_n=(2n+1)\pi\,T$.
Eq.~\ref{eq:quarkDSE} depends on two external quantities, the gluon propagator $D_{\mu\nu}(p)$ and the quark-gluon vertex $\Gamma_\mu(k;p,q)$, which will be discussed in the following sections.

\subsection{Gluon propagator}
\label{sec:gluon_prop}

\begin{figure}[tb]
 \begin{center}
 \includegraphics[width=0.45\textwidth]{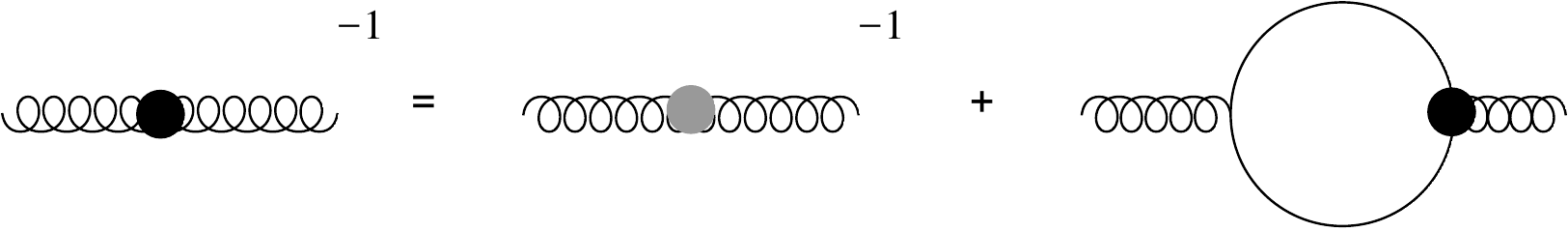}
 \caption{The gluon propagator DSE is split into a quenched part (gray blob) and the quark loop. The former is determined from quenched lattice results.}
 \label{fig:AADSEApprox}
 \end{center}
\end{figure}

The gluon propagator has its own DSE, which, however, is more complicated due to the appearance of two-loop terms and quadratic divergences. To obtain a quantitative description of the propagator but avoid these issues as far as possible, we apply the following approximation: We employ a fit to quenched lattice data for the gluon dressing functions and add unquenching effects via the quark loop, as is shown diagrammatically in \fref{fig:AADSEApprox}. This method was developed in a series of works. Initially, only the perturbative contributions from the quark propagators were taken into account \cite{Nickel:2006vf,Fischer:2009wc}. The most advanced variant includes the full quark propagator in the gluon propagator DSE and employs a model for the quark-gluon vertex \cite{Fischer:2012vc}.

This hybrid approach has the advantage that the full gluon propagator DSE does not need to be solved, but the full nonperturbative result from lattice calculations can be used. An equivalent solution from DSEs constitutes a considerable complication, since not only two-loop diagrams would need to be calculated to obtain a similar level of quantitative reliability, but also three- and four-point functions would need to be known at nonvanishing temperatures. To our knowledge, only some first results for these quantities are available in lattice \cite{Fister:2014bpa} and continuum approaches \cite{Huber:2013yqa,Huber:2016xbs}. Results for the propagators from continuum approaches were obtained, for example, by the functional renormalization group \cite{Fister:2011uw} and a perturbative analysis of a massive extension of Yang-Mills theory \cite{Reinosa:2013twa}.

\begin{figure*}[tb]
 \begin{center}
 \includegraphics[width=0.48\textwidth]{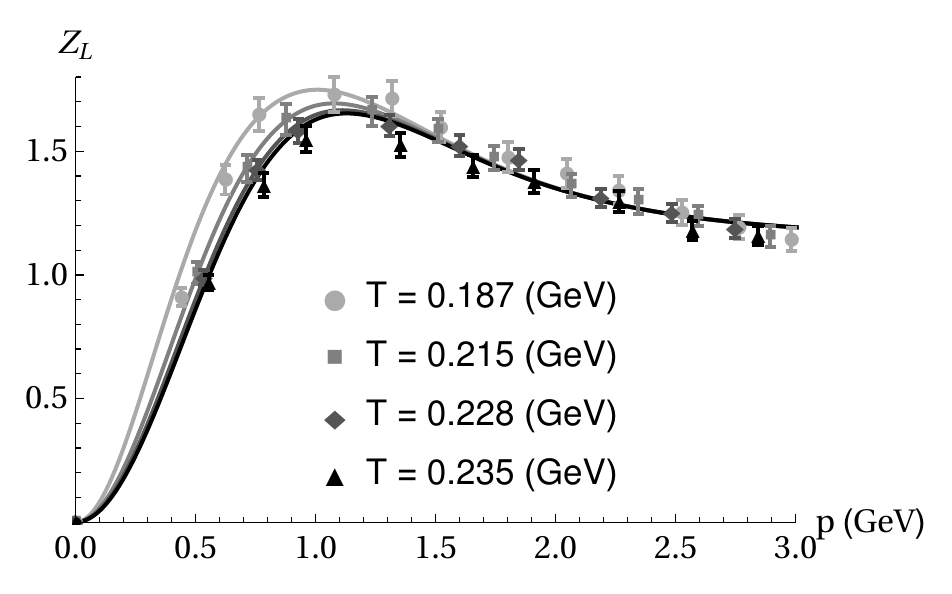}\hfill
 \includegraphics[width=0.48\textwidth]{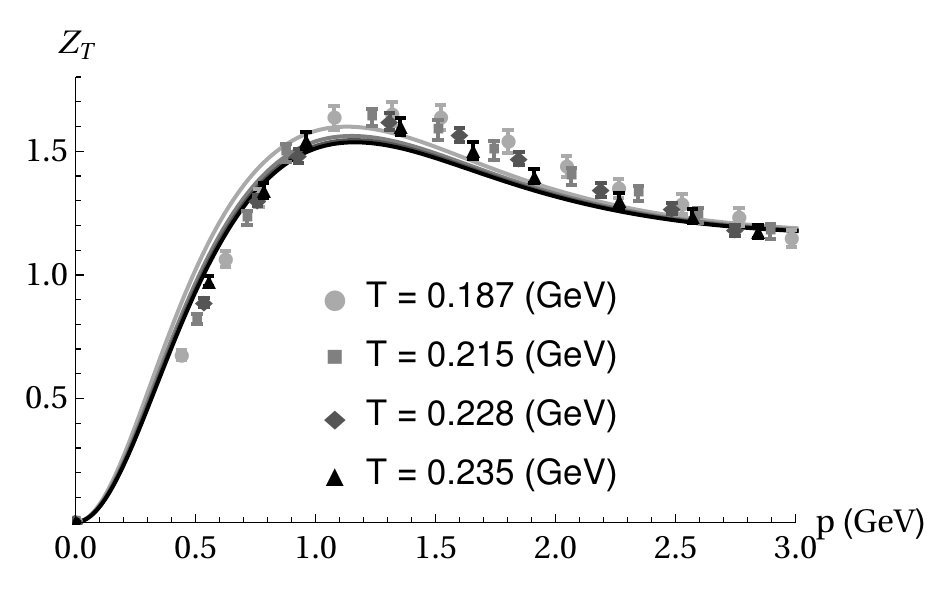}
 \caption{Chromoelectric (\textit{left}) and chromomagnetic (\textit{right}) unquenched gluon dressing functions for $SU(3)$ compared to unquenched lattice results corresponding to a pion mass of $m_{\pi} = 315 (\text{MeV})$ \cite{Aouane:2012bk}. For the comparison with lattice results, the renormalized quark mass is fixed via the Gell-Mann--Oakes--Renner relation as $m_R = 5.97\,\text{MeV}$ at $80\,\text{GeV}$. For the interaction strength $d_1=7\,\text{GeV}^2$ was used.}
 \label{fig:latZ}
 \end{center}
\end{figure*}

The drawback of this hybrid method is that no back-coupling effects of unquenching on the Yang-Mills sector can be taken into account. Nevertheless, the agreement with available lattice results when including only the direct effects via the quark loop is reasonable \cite{Fischer:2014ata}. As a check of our setup, we calculated the gluon propagator also for a higher quark mass to compare to lattice results, see \fref{fig:latZ}.\footnote{This calculation is similar to Ref.~\cite{Fischer:2014ata}, except that here we adjust the interaction strength parameter $d_1$, see \eref{eq:qug}, such that the transition temperature matches that of the $N_f=2$ lattice calculations \cite{Burger:2011zc}. Furthermore, we use the Gell-Mann--Oakes--Renner relation to fix the quark mass, whereas in Ref.~\cite{Fischer:2014ata} the Bethe-Salpeter equation for the pion was solved.}. Although we could in principle use fits of unquenched lattice data for the gluon propagator, we will use the hybrid method to allow for extensions to nonvanishing chemical potential later. Furthermore, such data is not available for all gauge groups considered here.

The fit function for the gluon dressing functions is \cite{Fischer:2010fx}
\begin{align}\label{eq:ZTL}
Z_{ T/L }(p^2)=\frac { x }{ (x+1)^2 } \Bigg( \left( \frac { c/\Lambda^2 }{ x+a_{ T/L } }  \right) ^{ b_{ T/L } }+\nnnl
 x\left( \frac { \alpha (\mu )\beta _{ 0 } }{ 4\pi  } \textrm{ln}(x+1) \right) ^{ \gamma  } \Bigg),
\end{align}
where $x=p^2/\Lambda^2$. We only fit the lowest Matsubara frequency. Dressings at higher Matsubara frequencies are evaluated by $Z^{T/L}(\vec{p}^2, p_4^2)=Z^{T/L}(p^2=\vec{p}^2+p_4^2,0)$, which is a good approximation according to lattice results \cite{Fischer:2010fx}.
The subscripts $T$ and $L$ refer to the splitting of the transverse gluon propagator at nonzero temperature:
\begin{align}
 D_{\mu\nu}(p)=P^L_{\mu\nu}(p)\frac{Z_L(p^2)}{p^2} + P^T_{\mu\nu}(p)\frac{Z_T(p^2)}{p^2},
\end{align}
where $P^T_{\mu\nu}(p)$ and $P^L_{\mu\nu}(p)$ project transversely and longitudinally to the heat bath, respectively:
\begin{align}
 P^T_{\mu\nu}(p)&=(1-\delta_{\mu4})(1-\delta_{\nu4})\left(\delta_{\mu\nu}-\frac{p_\mu p_\nu}{\vec{p}^2}\right),\\
 P^L_{\mu\nu}(p)&=P_{\mu\nu}-P^T_{\mu\nu}(p),\\
 P_{\mu\nu}&=\delta_{\mu\nu}-\frac{p_\mu p_\nu}{p^2}.
\end{align}

Eq.~\ref{eq:ZTL} is used to parametrize the quenched gluon propagators. The parameters $c = 11.5\,\text{GeV}^2$ and $\Lambda = 1.4\,\text{GeV}$ are kept fix. $\alpha(\mu)=g^2/4\pi$, which is also the value used for the strong coupling in the calculations, is fixed for the fits.
$\gamma=(-13N_c+4N_f)/(22N_c-4N_f)$ and $\beta_0$ are the anomalous dimension of the gluon propagator and the lowest coefficient of the $\beta$-function, respectively. In our calculations we inherit the scale from the fits.

\begin{figure*}[tb]
 \includegraphics[width=0.48\textwidth]{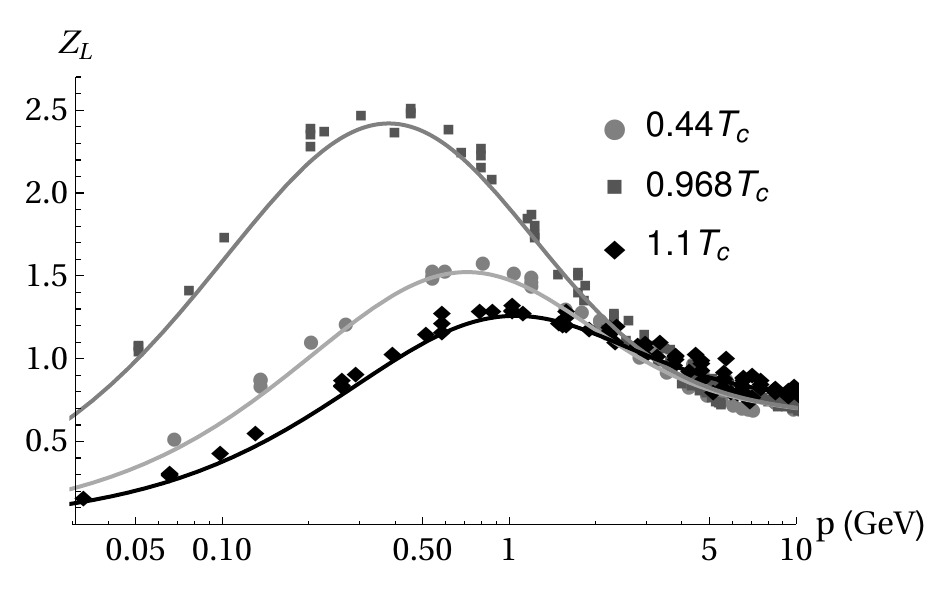}\hfill
 \includegraphics[width=0.48\textwidth]{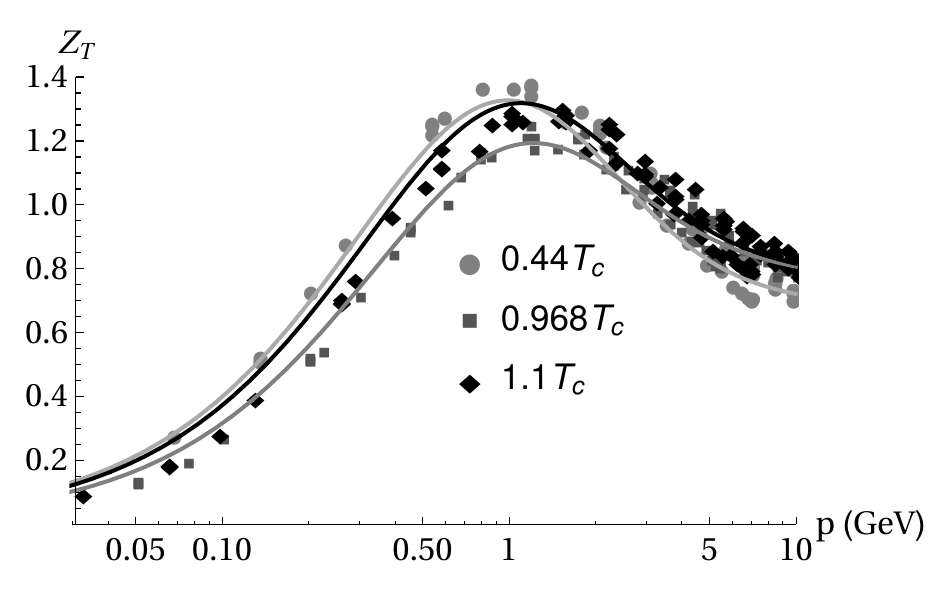}
 \caption{ Fitted quenched gluon dressing functions $Z_{L}$ (\textit{left}) and $Z_{T}$ (\textit{right}) for the gauge group $SU(2)$ compared to \cite{Maas:2011ez}.}
 \label{fig:ZTfit}
\end{figure*}

For the gauge group $SU(2)$ we obtain the fit parameters $a_{T/L}$ and $b_{T/L}$ from the lattice results of refs. \cite{Fischer:2010fx,Maas:2011ez}. Due to various uncertainties from the lattice input, the obtained values do not show a smooth behavior in temperature. To ameliorate that, the parameters themselves can be fitted. An equivalent procedure can be done for $SU(3)$ and the corresponding fit functions for can be found in \cite{Eichmann:2015kfa}. For $SU(2)$ the fit functions are given in Appendix~\ref{sec:fitsParameters}. The resulting dressing functions for $SU(2)$ are shown in \fref{fig:ZTfit} for selected temperatures.

For the gauge group $G_2$ we do not have any lattice data for the gluon propagator available. However, in two and three dimensions the quenched propagators were calculated in the vacuum \cite{Maas:2007af}. It was seen there  that qualitatively the gauge groups $SU(2)$, $SU(3)$ and $G_2$ behave the same and even quantitatively they are very similar. Of course, already in perturbation theory deviations are expected but they are subleading for the effects considered here. Since $G_2$-Yang-Mills theory has a first order transition, we use the $SU(3)$ results to construct an approximation for the $G_2$ gluon propagator: We use the $SU(3)$ fits for $a_{T/L}$ and $b_{T/L}$ with the corresponding value $\beta_0$ of $G_2$ and rescale the temperature to match the known critical temperature of $G_2$. The different $\beta_0$ is compensated in \eref{eq:ZTL} by a modification of the coupling constant at the renormalization point to maintain the form of the fits: $\alpha(\mu)=0.45$.
The parameters for the different gauge groups are summarized in \tref{tab:gaugeGroups}.

\begin{table}[b]
 \begin{center}
  \begin{tabular}{|c||c|c|c|c|c|c|c|}
  \hline
   & $C_A$ & $C_F$ & $\beta_0=\frac{11C_A}{3}$ & $c\,[GeV^2]$ & $\alpha(\mu)$ & $T^\text{qu}_c\,[MeV]$ \\
   \hline\hline
   $SU(3)$ & $3$ & $\frac{4}{3}$ & $11$ & $11.5$ & $0.3$ & $277\,\text{MeV}$ \\
   \hline
   $SU(2)$ & $2$ & $\frac{3}{4}$ & $\frac{22}{3}$ & $11.5$ & $0.3$ & $303\,\text{MeV}$ \\
   \hline
   $G_2$ & $2$ & $1$ & $\frac{22}{3}$ & $11.5$ & $0.45$ & $255\,\text{MeV}$ \\
   \hline
  \end{tabular}
 \caption{Differences between the gauge groups. Transition temperatures from  \cite{Fischer:2010fx} for $SU(2)$ and $SU(3)$ and from \cite{Cossu:2007dk,Ilgenfritz:2012wg} for $G_2$.}
 \label{tab:gaugeGroups}
 \end{center}
\end{table}

\subsection{Quark-gluon vertex}
\label{sec:qug}

Information on the quark-gluon vertex at nonvanishing temperatures is much scarcer than for the other correlation functions considered here. Thus, in the following we will rely on a model. 

The model employed in the following is given by \cite{Fischer:2009wc}
\begin{align}
\label{eq:qug}
&\Gamma_{\nu}(q;p,l) = \gamma_{\mu} \Gamma_{mod}(x) \nnnl
&\quad\times\Bigg(\frac{A(p^2) + A(l^2)}{2} \delta_{\mu, i}+\frac{C(p^2) + C(l^2)}{2} \delta_{\mu, 4} \Bigg),\\
&\Gamma_{mod}(x) = \frac{d_{1}}{\left(x+d_2\right)} \nnnl
&\quad+ \frac{x}{\Lambda^2 + x} \left(\frac{\alpha(\mu)\beta_{0}}{4 \pi}\textrm{ln}\left(\frac{x}{\Lambda^2} + 1\right)\right)^{2 \delta}.
 \label{eq:qglvertModel}
\end{align}
$p$ and $l$ are the antiquark and quark momenta, respectively, and $q$ is the gluon momentum. To guarantee multiplicative renormalizability, the choice for $x$ depends on the equation in which the vertex model is used. In the gluon propagator DSE, it is $(p^2+l^2)$ and in the quark propagator DSE $q^2$. $\Lambda$ and $\alpha(\mu)$ are chosen as for the gluon propagator fit. $\delta$ is the anomalous dimension of the ghost given by $\de=-9N_c/(44N_c-8N_f)$. The model is constructed such that the quark and the gluon propagator DSEs yield the correct anomalous dimensions. However, the model itself has twice the anomalous dimension of the quark-gluon vertex. This renormalization group motivated modification is necessary due to missing higher perturbative contributions.

The tensor structure of the model is restricted to the tree-level tensor of the vertex. Since it is known that the other seven transverse dressing functions are not negligible \cite{Hopfer:2013np,Williams:2014iea,Mitter:2014wpa,Williams:2015cvx,Aguilar:2016lbe}, their contributions must be effectively captured in the nonperturbative part of the vertex dressing which is characterized by the parameters $d_1$ and $d_2$. We fix $d_2=0.5\,\text{GeV}$ and discuss the determination of $d_1$ below. The resulting values for $d_1$ are summarized in \tref{tab:parameter}. The dependence on the quark dressing functions stems from the tree-level part of the Ball-Chiu construction \cite{Ball:1980ay} generalized to nonvanishing temperature.

\subsection{Derived quantities}

The observables to determine the transition temperatures will be the chiral condensate and the dual chiral condensate and suitable derivatives thereof. For a self-contained presentation we repeat the definitions of these quantities.

The chiral condensate is calculated as
\begin{align} \label{eq:chiralCondensate}
&\left<  \overline{\psi} \psi \right>  = -C_A Z_2 Z_m T \sum_{q_4}{\int{\frac{d^3q}{(2\pi)^3}\textrm{Tr}[S(q)]}}\nnnl
&\,=-4C_A Z_2 Z_m T \sum_{q_4}{\int{\frac{d^3q}{(2\pi)^3}\frac{B(q)}{A^2(q)\vec{q}^2 + C^2(q)q_4^2 + B^2(q)}}}.
\end{align}
We recall that a nonzero value means that chiral symmetry is broken. Although nonzero renormalized quark masses explicitly break chiral symmetry, its value at high temperatures is still considerably smaller than at small temperatures so that it can be used as an order parameter. The condensate is UV divergent. It is renormalized by subtracting a quark condensate with a heavier renormalized mass $m_s$ from a condensate with a light renormalized mass $m_l$:
\begin{align}
\Delta_{l,h} = -\left<  \overline{\psi} \psi \right>_{l} + \frac{m_l}{m_s}\left<  \overline{\psi} \psi \right>_{h}.
\end{align}

For the confinement/deconfinement transition we study the dual chiral condensate which is related to the Polyakov loop \cite{Bilgici:2008qy,Synatschke:2007bz} and has the same qualitative behavior, viz., it vanishes in the limit of infinitely heavy quark masses in the confined phase and obtains a nonzero value in the deconfinement phase. For finite quark masses the value at low temperatures is still very small and allows us to distinguish the two phases.

To compute the dual quark condensate $\Sigma$, we introduce generalized $U(1)$ valued boundary conditions for the quarks $\psi(x, 1/T) = e^{i \varphi } \psi(x, 0)$ where the physical condition is given by $\varphi = \pi$ \cite{Fischer:2009wc}. The dual quark condensate projects out the loops with winding number $1$. It is thus also called dressed Polyakov loop as it contains all loops winding around the time direction exactly once:
\begin{align}
 \Sigma=\int_0^{2\pi}\frac{d\varphi}{2\pi}e^{-i\,\varphi}\left<  \overline{\psi} \psi \right>_\varphi d\varphi.
\end{align}
The chiral condensate with the boundary angle $\varphi$ is calculated with \eref{eq:chiralCondensate} using generalized Matsubara frequencies depending on the boundary angle: $\omega'_n=(2n+\varphi/\pi)\pi\,T$.

Since QCD does not have a phase transition at vanishing chemical potential but a crossover, various transition temperatures can be defined. This needs to be kept in mind when comparing results. We use here the maxima of the following derivatives:
\begin{align}
 \chi_\text{ch} &= \frac{\partial \Delta_{l,h}}{\partial T},\\
 \chi_\text{dec} &= \frac{\partial \Sigma}{\partial T}.
\end{align}


\section{Results}
\label{sec:results}

\begin{table}[b]
 \begin{center}
  \begin{tabular}{|c|c|c|c|c|c|c|c|}
  \hline
   $N_f$ = 0 & $\alpha(\mu)$  & $d_1 (\text{GeV}^2)$ & &$N_f$ = 2 &$\alpha(\mu)$  & $d_1 (\text{GeV}^2)$  \\
   \hline\hline
   $SU(3)$ & 0.3 & $ 4.5 $&&$SU(3)$ & 0.3 & 7\\
   \hline
   $SU(2)$  & 0.3 & $ 7.3 $&&$SU(2)$ & 0.3 & 15\\
   \hline
   $G_2$  & 0.45 & $ 3.62 $&&$G_2$ & 0.45 & 6.78\\
   \hline
  \end{tabular}
 \caption{Parameters used for the quenched and $N_f=2$ computations.}
 \label{tab:parameter}
 \end{center}
\end{table}

\subsection{Quenched results}

\begin{figure*}[tb]
 \includegraphics[width=0.48\textwidth]{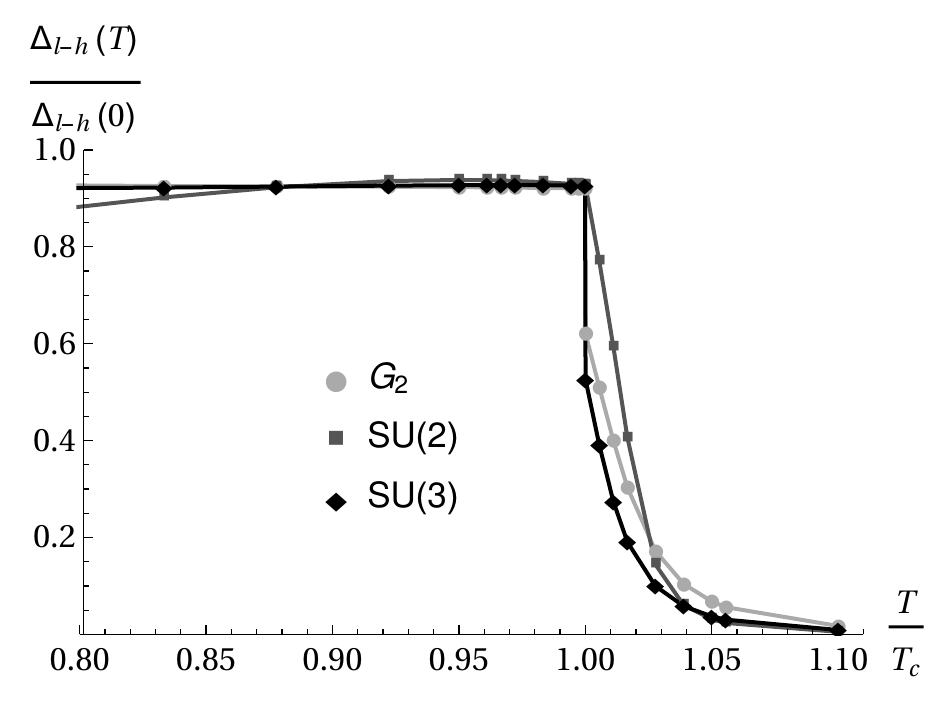}\hfill
 \includegraphics[width=0.48\textwidth]{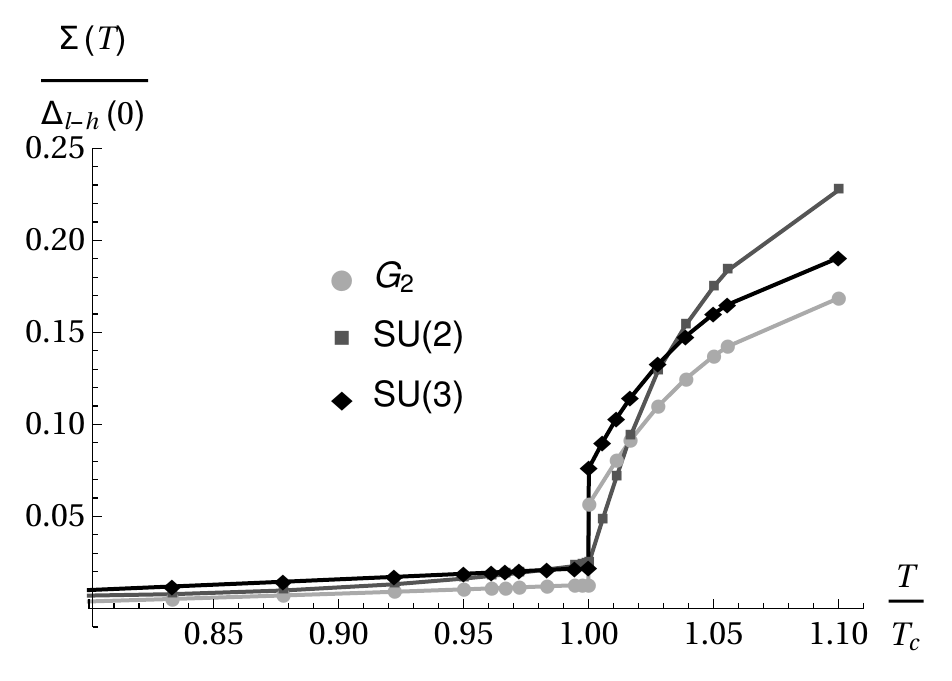}
 \caption{Chiral (\textit{left}) and dual chiral (\textit{right}) condensates normalized to the vacuum chiral condensates for the quenched theories.}
 \label{fig:ChiralDualnf0fig}
\end{figure*}

We first consider the quenched case. For this calculation, the renormalized quark mass is set to $m_R=3\,\text{MeV}$ at $80\,\text{GeV}$ as in \cite{Luecker:2013th} for all gauge groups. Results for the chiral and dual chiral condensates are show in \fref{fig:ChiralDualnf0fig}.

The results for both $SU(3)$ and $G_2$ are compatible with a first order phase transition, while those for $SU(2)$ indicate a second order transition.
The small decrease of the $SU(2)$ chiral condensate at the lower end of the shown temperature interval could be an artifact of the employed fit for the gluon dressing function. We explicitly checked that for low temperatures the chiral condensate stays close to the vacuum chiral condensate.
The nonzero dual chiral condensate below the critical temperature, which was already observed in earlier works, e.g., \cite{Fischer:2009wc,Luecker:2013th}, is due to the sensitivity of the dual chiral condensate to the quark-gluon vertex model: For different values of $d_1$ the deviation from zero varies.

Various schemes have been tested in the literature to set the values of $d_1$ and the quark masses, ranging from tuning them to lattice results for the chiral condensate to adjusting them to reproduce physical quantities like pion mass and decay constant. Here, we choose the value for $d_1$ in $SU(3)$ as in \cite{Luecker:2013th}, which is close to the value used in \cite{Fischer:2010fx}. For $SU(2)$, the value for $d_1$ was taken such as to reproduce the dual chiral condensate below the critical temperature with a behavior similar to that of $SU(3)$. The value of $d_1$ for $G_2$ was determined such as to obtain the $SU(3)$ vacuum chiral condensate times $2/3$. Although this is motivated by the overall color factor in the chiral condensate, in general this relation is not expected to hold and was taken due to the lack of any clear observable to use in the case of $G_2$. The values of $d_1$ are summarized in \tref{tab:parameter}.

The positions of the phase transitions are identical to the ones from the lattice input, see \tref{tab:gaugeGroups}, and determined directly by the behavior of the gluon propagators. The orders of the phase transitions, on the other hand, depend on the IR strength in the quark-gluon vertex model. By changing $d_1$ drastically, the order can be changed. In general, however, results look similar for small variations of $d_1$. The highest sensitivity we found for $SU(2)$ where we also saw some sensitivity to the detailed form of the gluon propagator fits. This and the behavior of the chiral condensate below the critical temperature mentioned above require a more detailed analysis in the future. For now, the nontrivial finding is that the expected behavior at both the chiral and deconfinement phase transitions can be reproduced with the same vertex model parameters.


\subsection{Unquenched results}

\begin{figure}[tb]
 \includegraphics[width=0.48\textwidth]{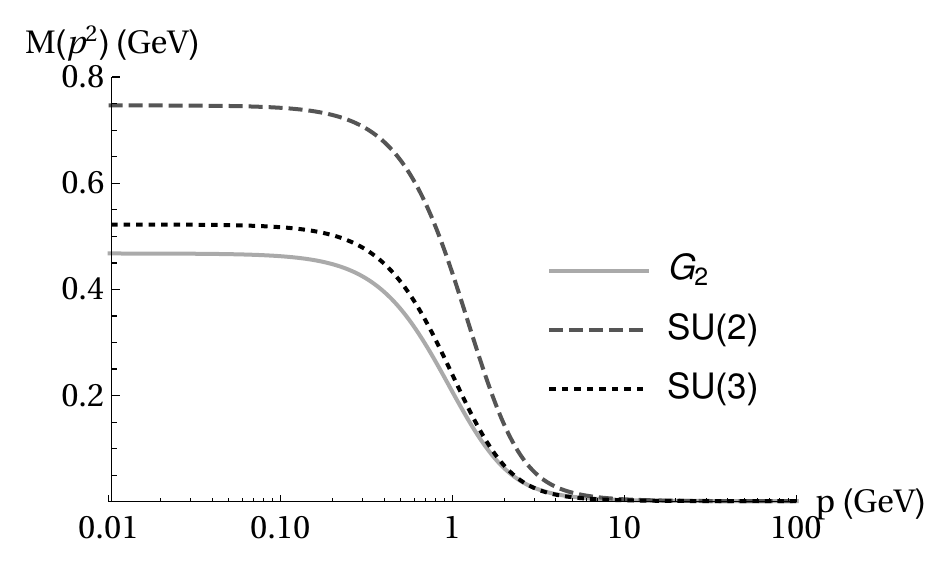}
 \caption{The quark mass function $M(p^2)=B(p^2)/A(p^2)$ as a function of momentum in the vacuum.}
 \label{fig:quark_mass}
\end{figure}

\begin{figure*}[tb]
 \includegraphics[width=0.48\textwidth]{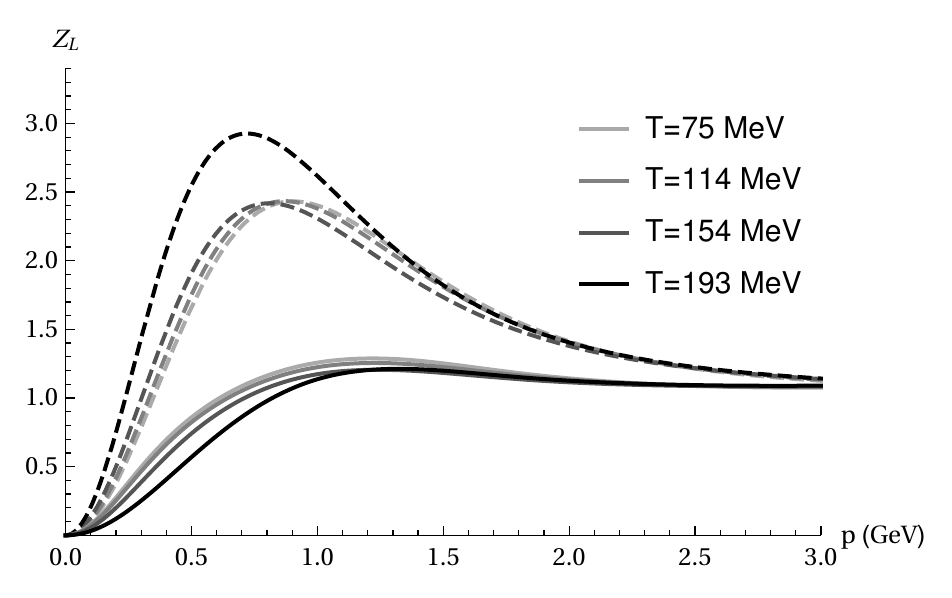}\hfill
 \includegraphics[width=0.48\textwidth]{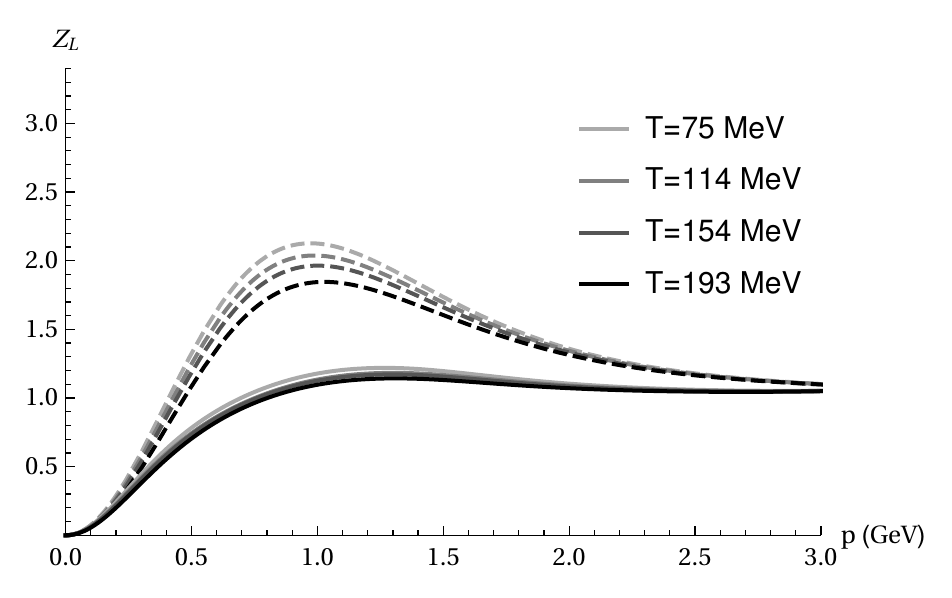}
 \caption{Chromoelectric (\textit{left}) and chromomagnetic (\textit{right}) gluon dressing functions for $SU(2)$. The dashed lines represent the quenched dressing functions.}
 \label{fig:Nf2SU2GluonDressing}
\end{figure*}

\begin{figure*}[tb]
 \includegraphics[width=0.48\textwidth]{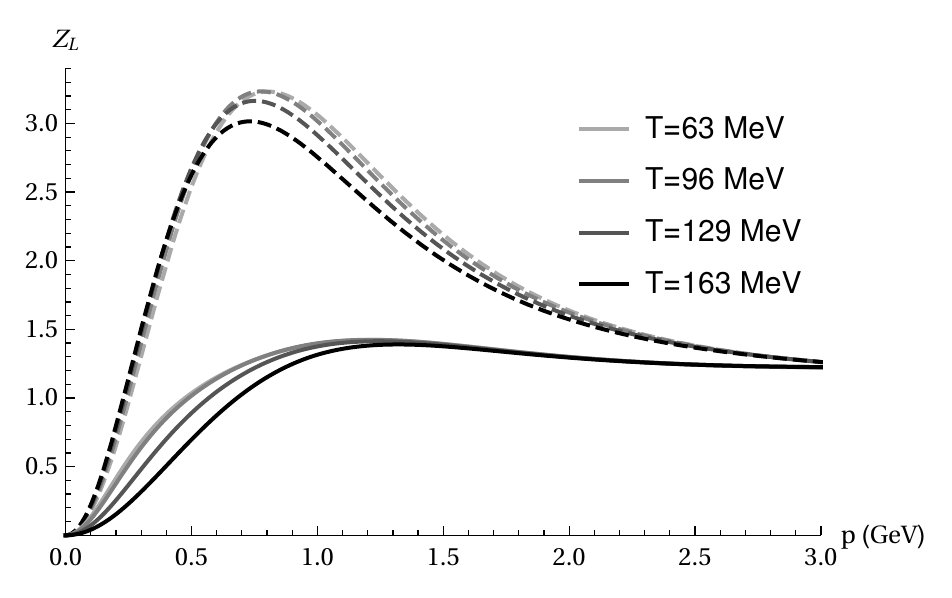}\hfill
 \includegraphics[width=0.48\textwidth]{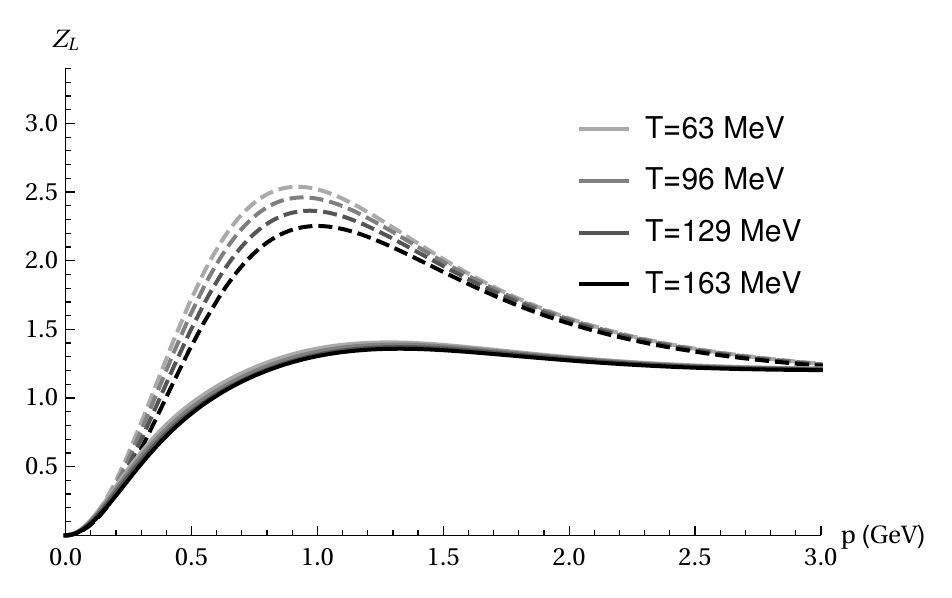}
 \caption{Chromoelectric (\textit{left}) and chromomagnetic (\textit{right}) gluon dressing functions for $G_2$. The dashed lines represent the quenched dressing functions.}
 \label{fig:Nf2G2GluonDressing}
\end{figure*}

The system is unquenched following the procedure described in Sec.~\ref{sec:gluon_prop}. We consider two light flavors with a renormalized mass of $m_R=1.18\,\text{MeV}$. The renormalization point is chosen as $80\,\text{GeV}$. This ensures that we are in the perturbative regime. The value of the quark mass at other points can be inferred from the mass function $M(p^2)=B(p^2)/A(p^2)$ shown in \fref{fig:quark_mass} for the vacuum. From the Gell-Mann--Oakes--Renner relation we obtain a pion mass of $140.9\,\text{MeV}$ corresponding to this choice of the quark mass. The effect of unquenching the gluon propagators is shown for $SU(2)$ and $G_2$ in \fref{fig:Nf2SU2GluonDressing} and \fref{fig:Nf2G2GluonDressing}, respectively. For $SU(3)$, the unquenched gluon dressing functions are compared to lattice results in \fref{fig:latZ}, where we adapted the quark mass to $m_R=5.97\,\text{MeV}$ to match the pion mass used in the lattice calculations via the Gell-Mann--Oakes--Renner relation. It should be noted that the chosen value for $d_1$ produces a transition temperature of $199\,\MeV$ which is close to the value of $202\,\MeV$ calculated on the lattice, but using a different definition of the transition temperature \cite{Burger:2011zc}.

\begin{figure*}[tb]
 \includegraphics[width=0.48\textwidth]{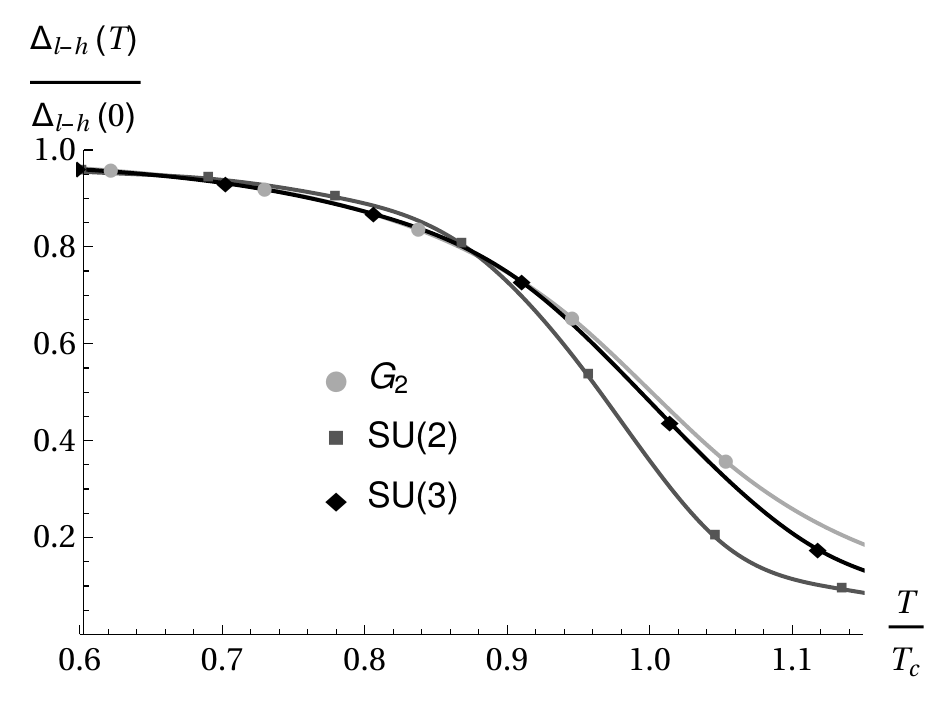}\hfill
 \includegraphics[width=0.48\textwidth]{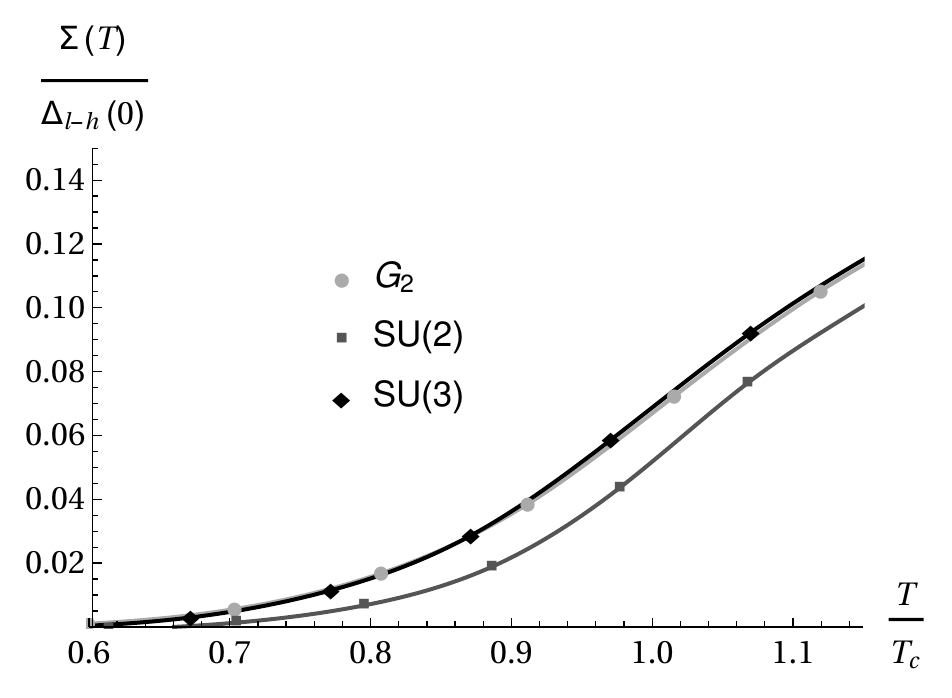}
 \caption{Chiral (\textit{left}) and dual chiral (\textit{right}) condensates normalized to the vacuum chiral condensates for $N_f = 2$.}
 \label{fig:Chidunf2fig}
\end{figure*}

As expected, the phase transitions become now crossovers as can be seen in \fref{fig:Chidunf2fig}. The resulting crossover temperatures are summarized in \tref{tab:temperaturesUnqu}.
The confinement/deconfinement crossover temperatures for $SU(3)$, $SU(2)$ and $G_2$ lie between $27\%$ and $40\%$ below the critical temperatures of the quenched systems.
The chiral crossover temperatures are close to the confinement/deconfinement crossover temperatures.

\begin{table}[b]
\begin{tabular}{|l||l|l|l|}
 \hline
 & $SU(3)$ & $SU(2)$ & $G_2$   \\\hline \hline
 $T_c$ (chiral) & 174 MeV & 218 MeV & 155 MeV \\
 \hline
 $T_c$ (deconfinement) & 182 MeV & 222 MeV & 160 MeV \\
 \hline
\end{tabular}
\caption{The crossover temperatures for the unquenched computations.}
\label{tab:temperaturesUnqu}
\end{table}

As explained in Sec.~\ref{sec:qug}, the employed quark-gluon vertex model encodes all nonperturbative information in the dressing function of the tree-level tensor, which has to be considered an effective dressing including also the dependence on the quark mass. Furthermore, effects of mesons and baryons are subsumed in the model. While the parameter $d_1$ can be adjusted, for example, such as to reproduce masses and decay constants of low-lying mesons in the vacuum \cite{Fischer:2014ata}, we are here interested in the situation at nonvanishing temperature. Since the model has only limited temperature dependence via the quark dressing functions, the value for $d_1$ describing the correct vacuum physics leads to a shifted value for the transition temperature \cite{Fischer:2014ata}.
Here, we first fix the value of $d_1$ to reproduce a transition temperature in the range of the $N_f=2$ lattice results for a pion mass of $315\,\MeV$. Then, we change the quark mass to obtain a pion mass of $140\,\MeV$. The resulting transition temperature is within the extrapolated interval given in Ref.~\cite{Burger:2011zc}. Note that the value for $d_1$ differs from those used elsewhere which were fixed in an $N_f=2+1$ calculation \cite{Luecker:2013th,Fischer:2014ata}.
For $SU(2)$, we tune $d_1$ to be in agreement with the deconfinement transition of $T_c = 217(23)$ from \cite{Cotter:2012mb}. However, it should be noted that we use lower quark masses than in the lattice reference. We have checked explicitly that for fixed $d_1$ increasing the renormalized quark mass increases also the transition temperature. The approach we take here is instead of tuning $d_1$ \textit{and} the quark mass for each gauge group, we take a similar value for the quark masses of all three theories, $m_R=1.2\,\text{MeV}$ in the case of $SU(2)$ and $m_R=1.18\,\text{MeV}$ for $SU(3)$ and $G_2$, which is set by the pion mass in $SU(3)$. To fix $d_1$ for $G_2$, we follow the same approach as in the quenched case. Again, the values of $d_1$ can be found in \tref{tab:parameter}.

\begin{figure}[tb]
 \begin{center}
 \includegraphics[width=0.45\textwidth]{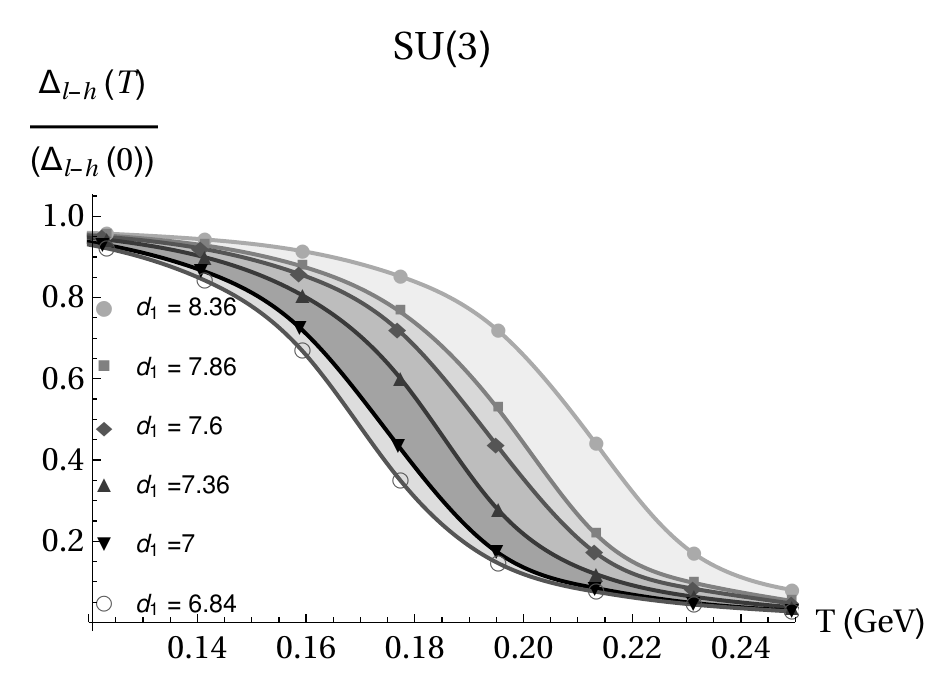}
 \caption{$SU(3)$ chiral condensate for variations of the quark-gluon vertex model.}
 \label{fig:chiralCondensatevaria}
 \end{center}
\end{figure}

To quantify the magnitude of the model parameter dependence, we show the effect of varying the parameter $d_1$ in the model, see \eref{eq:qglvertModel}, on the chiral condensate of $SU(3)$ in \fref{fig:chiralCondensatevaria}; see also \cite{Welzbacher:2016rcv}. The corresponding crossover temperatures for $SU(3)$, summarized in \tref{tab:chirald1varia}, vary by roughly the same amount as in the $N_f=2+1$ study of Ref.~\cite{Fischer:2014ata}, where different setups to determine $d_1$ where employed. In particular, one could take the value of $d_1$ to reproduce meson properties in the vacuum as a natural bound on $d_1$. In \fref{fig:chiralCondensatevaria}, this corresponds to the lower curve, whereas the central curve corresponds to the value of $d_1$ that reproduces the chiral condensate optimally around the transition. This is the value employed in the rest of this paper.
For $SU(2)$ and $G_2$, the dependence on $d_1$ is the same. This indication of universality of the employed truncation scheme - at least for vanishing chemical potential - is one of the main results of this work.

\begin{table}[b]
\begin{tabular}{|l||r|r|r|r|r|r|}
 \hline
  $d_1 $ (GeV$^2$) & 8.36 & 7.86 & 7.6 & 7.36 & 7 & 6.84   \\\hline 
 $T_c $ (MeV) & 212 & 201 & 194 & 185 & 174 & 169  \\\hline 
\end{tabular}
\caption{The chiral crossover temperatures for $SU(3)$, $N_f=2$ with $m_R\approx 1.2\,\text{Mev}$ as a function of $d_1$.}
\label{tab:chirald1varia}
\end{table}

\section{Summary}
\label{sec:summary}

We studied the universality of a DSE truncation scheme that relies on a modeled quark-gluon vertex and quenched lattice input for the gluon propagator. Unquenching was implemented by a hybrid approach that adds quark effects via the quark loop to the quenched gluon propagator data. In QCD, this truncation is well studied. Here we applied it to QC$_2$D and $G_2$-QCD. These theories can be studied with lattice methods at nonzero chemical potential where they can provide a test bed for truncations of functional equations. At vanishing chemical potential, all three theories are quite well understood. For this reason, we concentrated on this case for now to establish the usefulness of our approach. We found that the employed setup behaves universally and is able to reproduce the expected behavior of the confinement/deconfinement and chiral transitions for all three theories. This is promising for extensions to nonvanishing quark chemical potential which we plan to do as a next step \cite{Contant:2017onc}.

The employed approach also has some shortcomings. In particular, there is some sensitivity to the interaction strength of the vertex model. The nontrivial result is that solutions could be obtained for which both transitions are close to each other and that all three theories behave very similar under changes in the setup. If the latter feature persists at nonvanishing density, the development of further truncations of functional equations in QCD can profit from benchmarks provided by lattice calculations in QCD-like theories at nonvanishing chemical potential. Of course, the specifics of each theory have to be taken into account, for example, the emergence of a diquark condensate in QC$_2$D.

Improvements of the employed setup include an explicit calculation of the Yang-Mills part and an improved input for the quark-gluon vertex. The former point is challenging insofar, as not only the corresponding propagator equations require knowledge about three- and four-point functions, but also two-loop diagrams would need to be included for a quantitative description. For the quark-gluon vertex a variety of model extensions could be thought of. In the long run, however, an explicit inclusion of this quantity will be advisable.

\section{Acknowledgments}
We thank Andr\'e Sternbeck and Axel Maas for useful discussions and providing lattice data. We are grateful to Christian Fischer for helpful discussions and a critical reading of the manuscript. RC thanks the Institut f\"ur Theoretische Physik of the Justus-Liebig-Universit\"at Gießen, where this work was finished, for its hospitality. HPC Clusters at the University of Graz were used for the numerical computations. The software programs and packages \textit{Mathematica} \cite{Wolfram:2004}, \textit{DoFun} \cite{Alkofer:2008nt,Huber:2011qr} and \textit{CrasyDSE} \cite{Huber:2011xc} were used for deriving and solving numerically the DSEs. Feynman diagrams were created with Jaxodraw \cite{Binosi:2003yf}. 
Support by the FWF (Austrian science fund) under Contract No. P27380-N27 and through the Doctoral Program ``Hadrons in Vacuum, Nuclei and Stars'', Contract W1203-N16, is gratefully acknowledged.

\appendix

\begin{widetext}

\section{Fits of fit parameters $a_{T/L}$ and $b_{T/L}$ for $SU(2)$}
\label{sec:fitsParameters}

Using the fit function \eref{eq:ZTL} and the lattice data from \cite{Fischer:2010fx,Maas:2011ez}, we calculated values for the parameters $a_{T/L}$ and $b_{T/L}$ in $SU(2)$. Note that an overall factor was taken into account in the fit to accommodate for the finite renormalization of the dressing functions, but this factor was not employed in the calculation. To obtain a smooth behavior of the propagator, these values were fitted themselves using a polynomial ansatz. The resulting fits are shown in \fref{fig:abTLfit}. In contrast to the $SU(3)$ fit, we distinguish three different regions for the temperature dependence. However, the splitting of the region below the critical temperature is for convenience only, as it allows the use of simple polynomial fits, and does not reflect any physical changes. We tested explicitly that in derived quantities like the chiral condensate only the splitting at the critical temperature is visible. Since quenched $SU(2)$ has a continuous phase transition, we enforced continuity in the fits as well. The agreement of the transverse and longitudinal values at zero temperature was taken as another condition. A few exemplary fits are shown together with corresponding lattice data in \fref{fig:ZTfit}.

The fits used for the parameters $a_{T/L}$ and $b_{T/L}$ in terms of $t = \frac{T}{T_c} $ are
\begin{align}
a_T &= \left \{
	\begin{array}{l@{}@{\hspace{1em}}lr}
		0.46 \ge t  :& 1.41 + 0.43 t \\
		1 \ge t \ge 0.46 :& 1.52 + 0.20 t \\
		t \ge 1 :& 3.60 - 1.88t
	\end{array} \right. \\
b_T&=\left\{
	\begin{array}{l@{}@{\hspace{1em}}lr}
		0.49 \ge t  :& 2.20 + 0.07 t \\
		1 \ge t \ge 0.49 :& 2.43 - 0.40 t \\
		t \ge 1 :& 2.32 - 0.29t
	\end{array} \right. \\
a_L&=\left\{
	\begin{array}{l@{}@{\hspace{1em}}lr}
		0.53 \ge t  :& 1.41 - 2.09 t \\
		1 \ge t \ge 0.53 :& 0.89 - 1.51 t + 0.77 t^2 \\
		t \ge 1 :& -8.16 + 8.31t
	\end{array} \right. \\
b_L&=\left\{
	\begin{array}{l@{}@{\hspace{1em}}lr}
		0.52 \ge t  :& 2.20 - 1.82t \\
		1 \ge t \ge 0.52 :& 1.22 + 0.10 t - 0.05 t ^ 2 \\
		t \ge 1 :& -1.48 + 2.75t
	\end{array} \right. 
\end{align}

\begin{figure*}[tb]
 \begin{center}
 \includegraphics[width=0.48\textwidth]{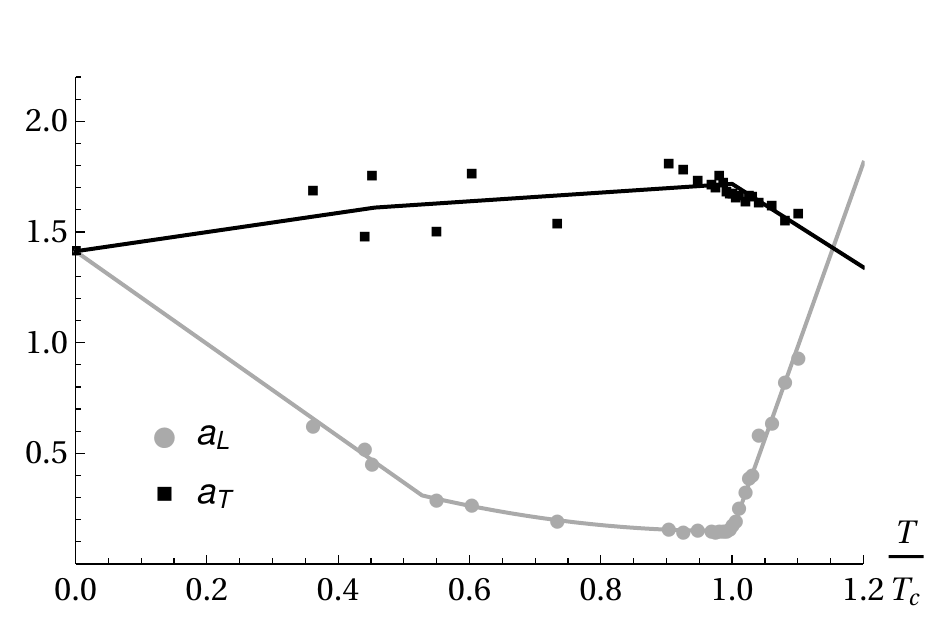}\hfill
 \includegraphics[width=0.48\textwidth]{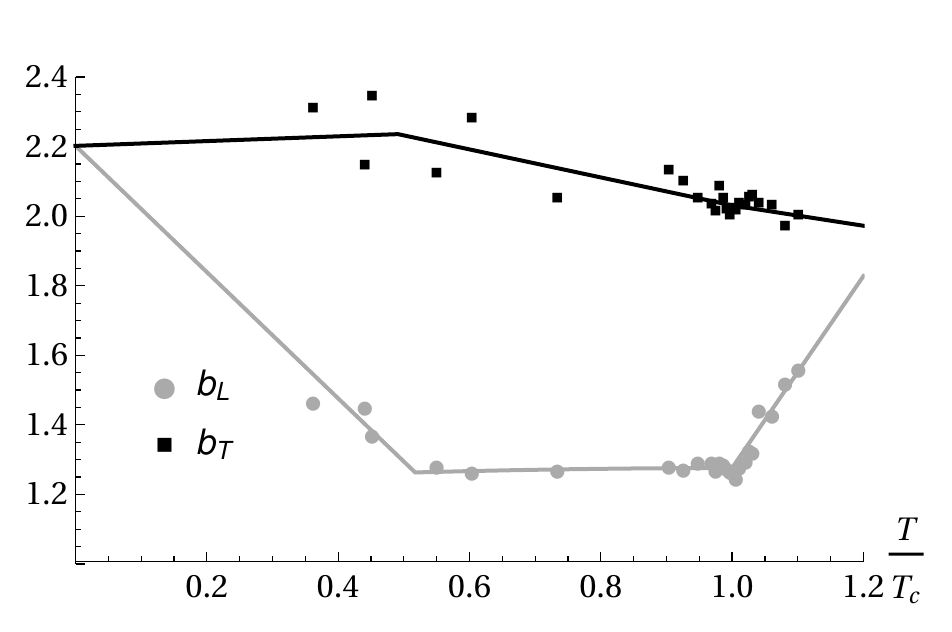}
 \caption{Parameters $a_{T/L}$ (black) and $b_{T/L}$ (gray) for fits of the gluon dressing functions of the gauge group $SU(2)$ .}
 \label{fig:abTLfit}
 \end{center}
\end{figure*}

\section{Quark dressing function $D(\vec p, \omega_n)$}
\label{sec:quarkD}

\begin{figure*}[tb]
 \includegraphics[width=0.48\textwidth]{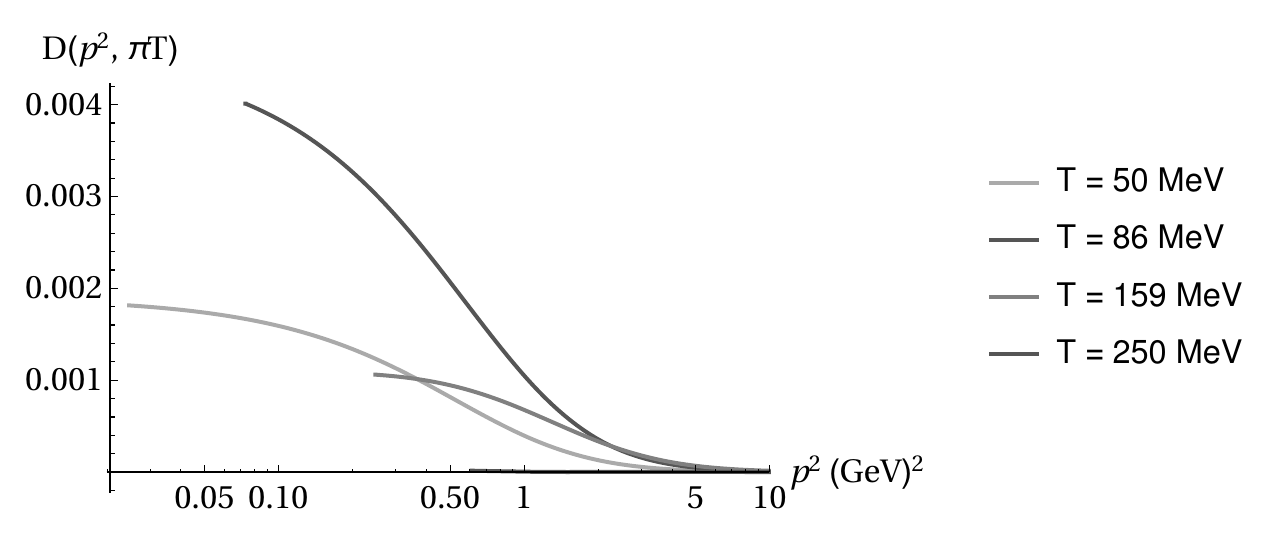}\hfill
 \includegraphics[width=0.48\textwidth]{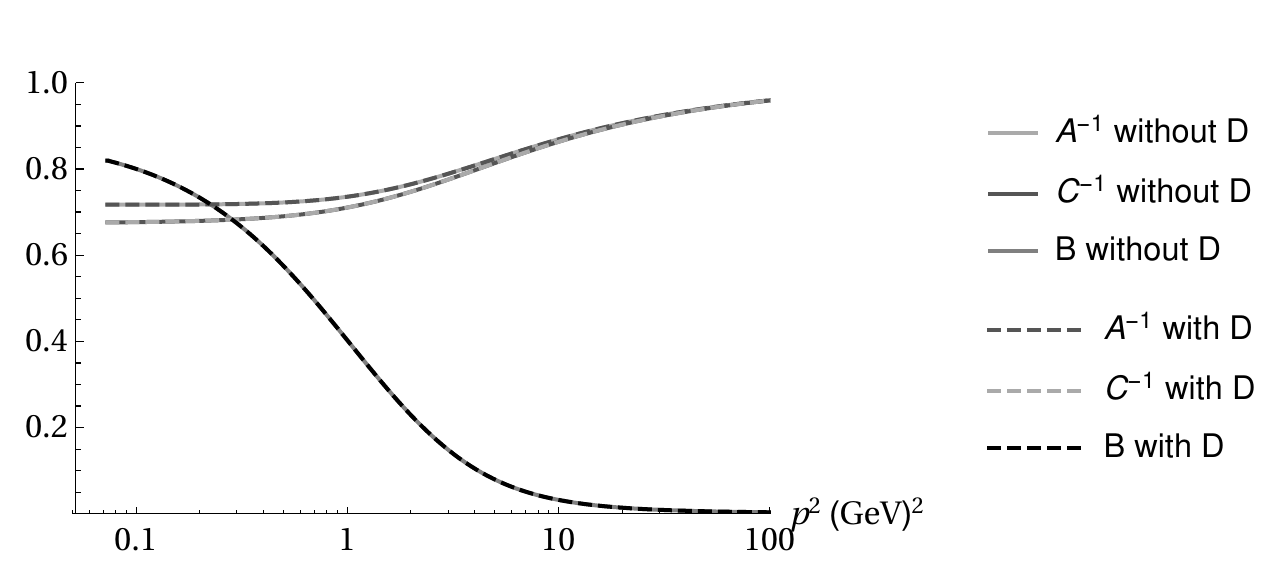}
 \caption{\textit{Left:} Evolution of the dressing function $D(\vec{p},\omega_1)$ for various temperatures.
 \textit{Right:} Effects of the dressing function $D(\vec{p},\omega_n)$ on $A(\vec{p},\omega_1)$, $B(\vec{p},\omega_1)$ and $C(\vec{p},\omega_1)$ at $T = 86\,\text{MeV}$.}
 \label{fig:Ddressing}
\end{figure*}

The dressing function $D(\vec{p},\omega_n)$ is typically neglected, since it is very small. We tested its influence on our calculations.
The left plot in \fref{fig:Ddressing} shows the evolution of the dressing function $D(\vec{p},\omega_n)$ for different temperatures and $N_f=2$. It is basically zero at very low temperatures, starts to increase until approximatively $T=100\,\text{MeV}$ and then decreases again. In the right plot of \fref{fig:Ddressing}, the effect on the other quark dressing functions is shown when including $D(\vec{p},\omega_n)$. The relative difference for the quark propagator between the computations with and without $D(\vec{p},\omega_n)$ function is below $0.0001$. Consequently, dropping this dressing in all our calculations is well justified.
\end{widetext}

\bibliographystyle{utphys_mod}
\bibliography{literature_quAtT}

\end{document}